\documentclass[twocolumn,superscriptaddress,aps,prb]{revtex4}
\usepackage{graphicx}
\usepackage[novolumeabbr]{unitsdef} 
\newunit{\invcm}{\centi\meter\unitsuperscript{-1}}

\usepackage{subfigure}

\begin{document}
\title{Coherent energy scale revealed by ultrafast dynamics of UX$_{3}$ (X=Al, Sn, Ga) single crystals}
\author{Saritha K. Nair}
\affiliation{Division of Physics and Applied Physics, School of Physical and Mathematical Sciences, Nanyang Technological University, Singapore 637371, Singapore}
\author{J.-X. Zhu}
\affiliation{Los Alamos National Laboratory, Los Alamos, New Mexico 87545, USA}
\author{J. L. Sarrao}
\affiliation{Los Alamos National Laboratory, Los Alamos, New Mexico 87545, USA}
\author{A. J. Taylor}
\affiliation{Los Alamos National Laboratory, Los Alamos, New Mexico 87545, USA}
\author{Elbert E. M. Chia}
\affiliation{Division of Physics and Applied Physics, School of Physical and Mathematical Sciences, Nanyang Technological University, Singapore 637371, Singapore}

\begin{abstract}
Temperature dependence of relaxation dynamics of UX$_{3}$ (X = Al, Ga, Sn) compounds is studied using time resolved pump-probe technique in the reflectance geometry. UGa$_{3}$ is an itinerant antiferromagnet, while UAl$_{3}$ and USn$_{3}$ are spin fluctuation systems. For UGa$_{3}$, our data are consistent with the formation of a spin density wave SDW gap as evidenced from the quasidivergence of the relaxation time $\tau$ near the N\'{e}el temperature $T_{N}$. For UAl$_{3}$ and USn$_{3}$, the relaxation dynamics shows a change from single exponential to two exponential behavior below a particular temperature, suggestive of coherence formation of the 5\textit{f} electrons with the conduction band electrons. This particular temperature can be attributed to the spin fluctuation temperature $T_{sf}$, a measure of the strength of Kondo coherence. Our $T_{sf}$ is consistent with other data such as resistivity and susceptibility measurements. The temperature dependence of the relaxation amplitude and time of UAl$_{3}$ and USn$_{3}$ were also fitted by the Rothwarf-Taylor model. Our results show ultrafast optical spectroscopy is sensitive to c-\textit{f} Kondo hybridization in the \textit{f}-electron systems. \end{abstract}

\date{\today}
\maketitle
\section{Introduction}
The uranium compounds UX$_{3}$, where X is a IIIb (Al, Ga, In, Tl) or IVb (Si, Ge, Sn, Pb) element, crystallize in the cubic AuCu$_{3}$-type structure~\cite{Fournier1985} and have U-U distances ($d_{U-U}$) much larger than the Hill limit ($\sim$~3.5~\AA) for uranium compounds.\cite{Hill1970} The different degree of hybridization of the 5\textit{f} electron orbitals with the conduction electron orbitals in these compounds leads to a wide range of magnetic behavior such as Pauli enhanced paramagnetism (UAl$_{3}$, USi$_{3}$, and UGe$_{3}$), antiferromagnetism (UGa$_{3}$, UPb$_{3}$ and UIn$_{3}$), and heavy fermion behavior (USn$_{3}$).\cite{Fournier1985,Koelling1985,Ott1985} Due to the the above-mentioned properties and the availability of high quality crystals, UX$_{3}$ compounds are ideal candidates for studying how physical properties and underlying electronic structure are related.

The anomalous behavior of the resistivities of UX$_{3}$ compounds can be explained on the basis of spin fluctuations in narrow 5\textit{f} bands.\cite{Arko1972,Jullien1976} A temperature characteristic of the spin fluctuations in the UX$_{3}$ compounds is the spin fluctuation temperature, $T_{sf}$, which expresses the strength of hybridization between \textit{f} and conduction electrons (c-\textit{f} hybridization). The degree of hybridization is related to the degree of delocalization of the \textit{f}-electrons. A high value of $T_{sf}$ corresponds to more easily hybridized (delocalized) electrons. Above $T_{sf}$, \textit{f}-electrons are localized; whereas below $T_{sf}$, there is quasiparticle coherence from the hybridization between \textit{f}-electrons and conduction electrons, \textit{i.e.}, \textit{f}-electrons now become more delocalized (or itinerant). The effective hybridization below $T_{sf}$ leads to changes in measured physical properties. For example, the electrical resistivity changes from a $T$-linear law above $T_{sf}$ to a $T$-quadratic law below this temperature.\cite{Buschow1972,Jullien1973,Jullien1974a,Jullien1976} The temperature at which the magnetic susceptibility reaches a Curie-Weiss law is theoretically of the order of $T_{sf}$.\cite{Jullien1976} A modified Curie-Weiss law, i.e. $\chi(T)=\chi_{0}+C/(T+T^{\ast})$, associates $T^{\ast}$ with $T_{sf}$ for relatively strong \textit{c}-\textit{f} hybridization.\cite{Lin1985,Yuen1990}

Ultrafast time-resolved pump-probe spectroscopy has been recognized as a powerful technique to study the nonequilibrium carrier dynamics in strongly correlated electron materials. In addition to distinguishing different phases in a material by their different relaxation dynamics, it can discern whether one phase coexists or competes with another phase in close proximity,\cite{Chia2007,Nair2010} giving information on the nature of low energy electronic structure of correlated electron systems, for example, in high-temperature superconductors. Pump-probe experiments have also been performed on actinide compounds, such as the itinerant antiferromagnets UNiGa$_{5}$ and UPtGa$_{5}$,~\cite{Chia2006,Chia2011} and the heavy-fermion superconductor PuCoGa$_{5}$.~\cite{Talbayev10}

The hybridization between the conduction electrons and the localized \textit{f} electrons also causes a narrow gap to form in the density of states near the Fermi level.~\cite{Demsar06b} This gap, called the hybridization gap, results in a relaxation bottleneck, evidenced by an increase in the relaxation time $\tau$ at low temperatures. For example, in heavy fermions such as YbAgCu$_{4}$ and SmB$_{6}$, $\tau$ increases monotonically with decreasing temperature.~\cite{Demsar06b}. The temperature dependence of the relaxation amplitude and time were fit using the Rothwarf-Taylor (RT) model. In this paper, we investigate the ultrafast dynamics in three isostructural uranium compounds, UAl$_{3}$, UGa$_{3}$ and USn$_{3}$, using the ultrafast pump-probe technique. The variation in hybridization strength is responsible for the differences in properties of these three isostructural compounds. UAl$_{3}$ and USn$_{3}$ are categorized as spin-fluctuation systems.\cite{Buschow1972,VanMaaren1974,Lupsa1994,Lupsa1996,Norman1986,Cornelius1999} UGa$_{3}$ does not behave as a spin fluctuation system, but is an itinerant 5\textit{f} electron antiferromagnet. In fitting the transient change in reflectivity for UAl$_{3}$ and USn$_{3}$, we needed a two-exponential decay function below $T_{sf}$, which points to the presence of two relaxation channels below $T_{sf}$. This arises from the hybridization between \textit{f} electrons and conduction electrons below $T_{sf}$. This shows that the ultrafast pump probe technique is sensitive to \textit{c}-\textit{f} hybridization in \textit{f}-electron systems. Our $T_{sf}$ is consistent with that obtained from resistivity and susceptibility measurements. We were also able to fit the temperature dependence of the relaxation amplitude and time using the RT model. For UGa$_{3}$, the relaxation time diverges as the temperature approaches the N\'{e}el temperature $T_{N}$, corresponding to the formation of a spin density wave (SDW) gap near the Fermi level.

\begin{figure}[hbt!]
\includegraphics[width=8cm]{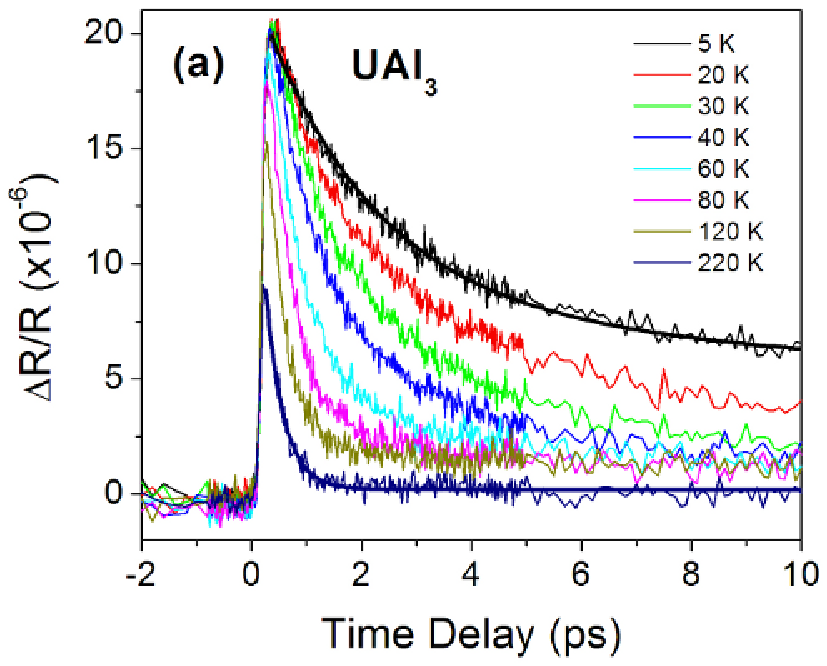}\\
\includegraphics[width=8cm]{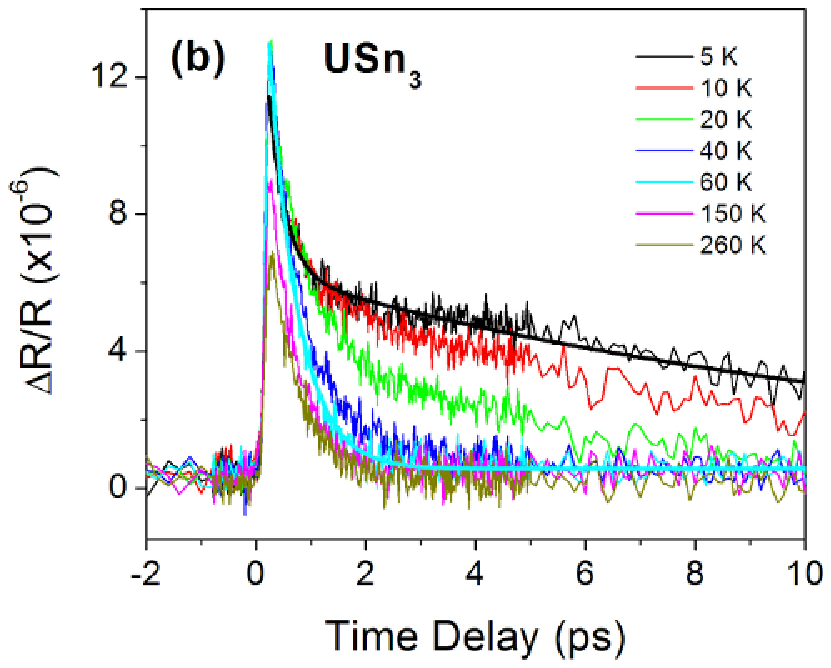}
\caption[]{Transient reflection ${\Delta}R/R$ versus pump-probe time delay at different temperatures for (a) UAl$_{3}$, and (b) USn$_{3}$. Thick solid curves denote exponential fits of data.}
\label{fig:UAl3USn3}
\end{figure}

\section{Experimental Setup}
In our pump-probe experimental setup in reflectance geometry, a Ti:sapphire laser producing sub-100~fs pulses at $\approx$800~nm (1.55~eV) was used as a source of both pump and probe pulses. The pump and probe pulses were cross polarized. The pump spot diameter was 60~${\mu}$m and that of probe was 30~${\mu}$m. The reflected probe beam was focused onto an avalanche photodiode detector. The photoinduced change in reflectivity (${\Delta}R/R$) was measured using lock-in detection. In order to minimize noise, the pump beam was modulated at 100~kHz with an acousto-optical modulator. The experiments were performed with an average pump power of 2 mW, giving a pump fluence of $\sim$1~$\mu$J/cm$^{2}$. The probe intensity was approximately ten times lower. Data were taken from 10~K to 300~K. The experiments were performed on single crystals of UX$_{3}$ (X = Al, Ga, Sn) grown using standard flux technique, with X used as the flux in each case.\cite{Cornelius1999}

\begin{figure}[h!]
\includegraphics[width=8cm]{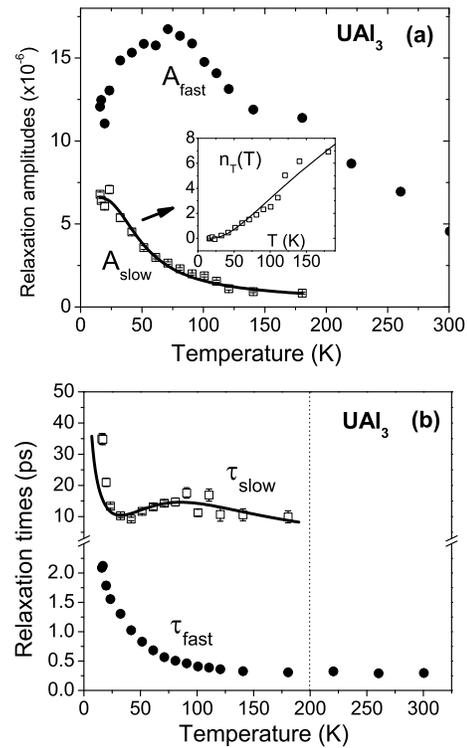}
\caption[]{Temperature dependence of (a) amplitudes and (b) relaxation times for UAl$_{3}$. Solid lines are fits to the RT model of the slow component.}
\label{fig:UAl3amptau}
\end{figure}

\begin{figure}[h!]
\includegraphics[width=8cm]{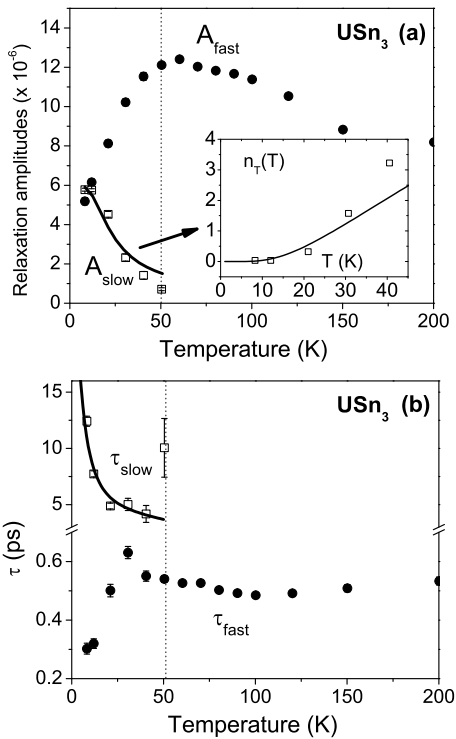}
\caption[]{Temperature dependence of (a) amplitude and (b) relaxation time for USn$_{3}$. Solid lines are fits to the RT model of the slow component.}
\label{fig:USn3amptau}
\end{figure}

\section{UA\lowercase{l}$_{3}$ and US\lowercase{n}$_{3}$}
In Fig.~\ref{fig:UAl3USn3} we show the ${\Delta}R/R$ at different temperatures for (a) UAl$_{3}$ and (b) USn$_{3}$, as a function of the time delay between the pump and probe pulses. In both UAl$_{3}$ and USn$_{3}$, only a fast relaxation of $\sim$500~fs, which is typical of regular metals, is observed at high temperatures. At low temperatures, an additional slow, positive picosecond relaxation is observed. Data at low temperatures are fitted to the two-exponential decay function $\Delta R/R (t)= A_{fast}(T) \exp (-t/\tau_{fast}) + A_{slow}(T) \exp(-t/\tau_{slow})$. This change from one- to two-exponential decay occurs at a particular crossover temperature --- $\sim$200~K for UAl$_{3}$ and $\sim$50~K for USn$_{3}$, suggestive of two relaxation channels below this crossover temperature. These crossover temperatures are of the order of the spin fluctuation temperatures $T_{sf}$ obtained in these compounds from temperature-dependent electrical resistivity and magnetic susceptibility measurements ($\sim$150~K for UAl$_{3}$~\cite{Aoki2000,Jullien1976} and $\sim$50~K for USn$_{3}$ \cite{Jullien1976,Loewenhaupt1990,Sugiyama2002}) We thus associate this crossover temperature to the spin fluctuation temperature $T_{sf}$.

\begin{figure}
\centering \includegraphics[width=8cm,clip]{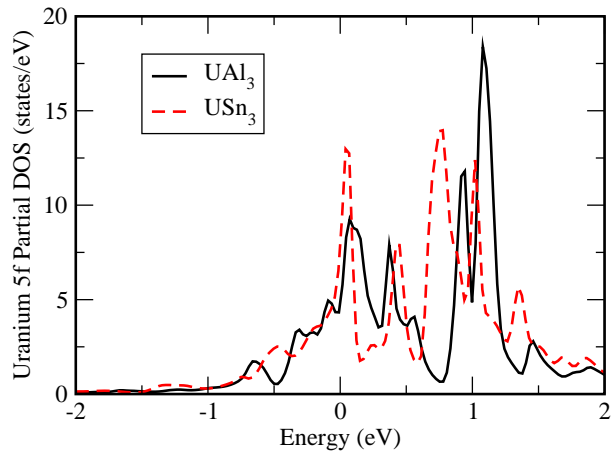}
\caption{Total DOS calculated from the LAPW method for UAl$_{3}$ and USn$_{3}$, in the magnetic unit cell, in the energy range (-2,2) eV. Note the narrower peak width near the Fermi energy ($E$=0) in USn$_{3}$ compared to UAl$_{3}$.}
\label{fig:UX3dos}
\end{figure}

To understand the different characteristic temperatures in UAl$_{3}$ and USn$_{3}$, we have also performed band structure calculations in the framework of the density functional theory, by using the WIEN2k linearized augmented plane wave method.\cite{Blaha01} A generalized gradient approximation \cite{Perdew1996} was used to treat exchange and correlation. Spin-orbit coupling was included in a second-variational way.  The obtained U partial 5f density of states, as shown in Fig.~\ref{fig:UX3dos}, indicates a narrower peak width near the Fermi energy, in USn$_{3}$ as compared with UAl$_{3}$. In addition, one can see that the splitting between the two major peaks is smaller in USn$_{3}$ than in UAl$_{3}$. In view of the fact that the spin-orbit coupling is quite local to the U atoms, one would expect the same effect on both USn$_{3}$ and UAl$_{3}$. A reasonable explanation for this difference is a smaller hybridization gap in USn$_{3}$ compared to UAl$_{3}$, due to the weakening of the hybridization in USn$_{3}$ --- a result of the lattice expansion ($a$=4.626 \AA~in USn$_{3}$ versus $a$=4.264 \AA~in UAl$_{3}$).\cite{Aoki2000} Though conventional band structure calculations underestimate the correlation effect, the trend of smaller coherence energy scale in USn$_{3}$ than in UAl$_{3}$ should be robust, as has recently been exemplified in other isostructural actinide compounds.\cite{Zhu2012}

In this context, the two-exponential behavior at low temperature can be explained by the \textit{c}-\textit{f} hybridization occurring below $T_{sf}$. Below $T_{sf}$, the interaction of partially-filled \textit{f} shell electrons with conduction electrons lead to the formation of heavy quasiparticles.\cite{Lobad2000} As the \textit{f}-electrons are localized above $T_{sf}$, relaxation occurs through phonon channel only. Hence only a single exponential decay is expected above $T_{sf}$. When $T<T_{sf}$, the spin fluctuation channel opens up due to hybridization. Electrons now relax via \textit{both} phonon and spin fluctuation channels resulting in a two-exponential decay behavior. Also, a higher $T_{sf}$ value in UAl$_{3}$ compared to USn$_{3}$ points to a stronger \textit{c}-\textit{f} hybridization, which is expected, as \textit{c}-\textit{f} hybridization tends to decrease as the size of the non-\textit{f} atom increases,~\cite{Arko1978,Koelling1985} which causes the lattice expansion as we discussed above.

The hybridization between the conduction band and the localized \textit{f}-levels also results in the formation of a narrow gap in the density of states near the Fermi level, called the hybridization gap. The presence of this gap causes a bottleneck in quasiparticle relaxation, resulting in a divergence of the relaxation time at low temperatures. The temperature dependence of the relaxation amplitude and relaxation time can be quantitively explained by the Rothwarf-Taylor (RT) model. It is a phenomenological model that was used to describe the relaxation of photoexcited superconductors, \cite{Rothwarf67,Demsar03c} itinerant antiferromagnets \cite{Chia2006,Chia2011} and heavy-fermion metals, \cite{Demsar06b} where the presence of a gap in the electronic density of states gives rise to a relaxation bottleneck for carrier relaxation. In heavy fermions, after the initial photo-excitation by a pump pulse, the subsequent fast relaxation due to electron-electron scattering results in excess densities of electron-hole pairs (EHPs) and high-frequency phonons (HFPs). When an EHP with energy $\geq \Delta$ ($\Delta$= hybridization gap) recombines, a HFP is created. The HFPs released in the EHP recombination are trapped within the excited volume and can re-excite EHPs; hence they act as a bottleneck for EHP recombination, and recovery is governed by the decay of the HFP population. The evolution of EHP and HFP populations is described by a set of two coupled nonlinear differential equations.

\begin{figure}
\includegraphics[width=8cm]{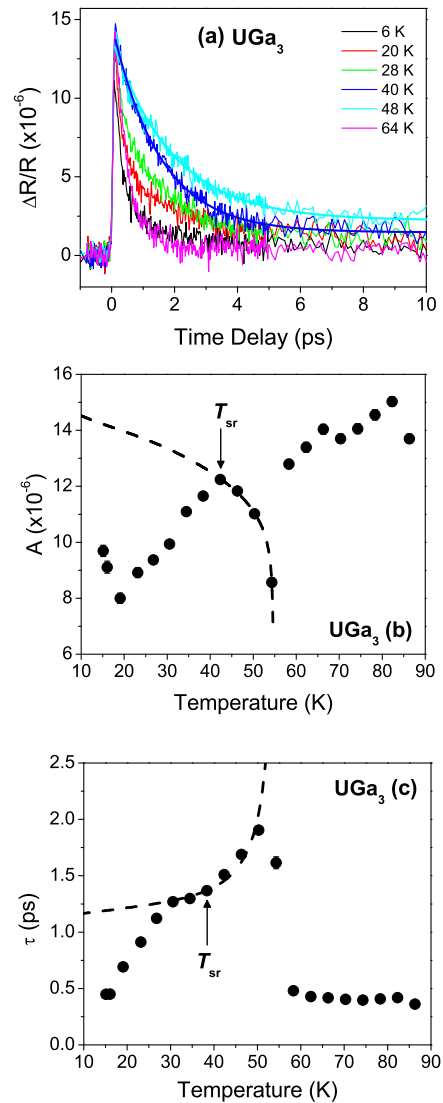}
\caption{(a) Photoinduced change in reflectivity ${\Delta}R/R$ versus pump-probe time delay at different temperatures of UGa$_{3}$. Thick blue (cyan) curves denote one-exponential fits of data at $T$=40~K (48~K). Temperature dependence of (b) amplitude and (c) relaxation time for UGa$_{3}$ obtained from one-exponential fits, with dashed lines in (b) and (c) being fits to the RT model from 40~K to $T_{N}$.}
\label{fig:UGa3}
\end{figure}

The results of the RT model are as follows:\cite{Kabanov05,Demsar06b} from the amplitude $A(T)$, one obtains the density of thermally excited EHPs $n_{T}$ via the relation
\begin{equation}
n_{T}(T) \propto  \mathcal{A}(T)^{-1} -1 \label{eqn:nTA}
\end{equation} where $\mathcal{A}(T)$ is the normalized amplitude
($\mathcal{A}(T)=A(T)/A(T \rightarrow 0)$). Then we fit the experimental $n_{T}(T)$ to the expression\cite{Demsar06b}
\begin{equation}
n_{T}(T) \propto  \sqrt{T} \exp (-\Delta/T),
\label{eqn:nTHF}
\end{equation} where the hybridization gap $\Delta$ is temperature independent (or very weakly temperature dependent) and can be obtained from the fitting. Moreover, for a constant pump intensity, the temperature-dependence of $n_{T}$ also governs the temperature-dependence of $\tau^{-1}$, given by
\begin{equation}
\tau^{-1}(T) =  \Gamma [\delta (\beta n_{T} +1)^{-1} + 2 n_{T}](\Delta + \alpha T \Delta^{4})
\label{eqn:RTtau}
\end{equation} where $\Gamma$, $\delta$, $\beta$ and $\alpha$ are $T$-independent fitting parameters.

Since, below $T_{sf}$, the second relaxation component appears, we attribute it to relaxation across the hybridization gap, and use the RT model to fit the its amplitude and relaxation time below $T_{sf}$ in UAl$_{3}$ and USn$_{3}$. The inset of Fig.~\ref{fig:UAl3amptau}(a) shows $n_{T}(T)$ obtained from $A_{slow}(T)$ using Eq.~(\ref{eqn:nTA}), with the solid line being the fit to Eq.~(\ref{eqn:nTHF}), with the fitting parameter $\Delta \approx (230 \pm 10)$~K. The fitted values of $n_{T}(T)$ are then inserted into Eq.~(\ref{eqn:RTtau}) to fit the experimental values of $\tau_{slow}(T)$, shown in Fig.~\ref{fig:UAl3amptau}(b). Similar fits are also done for USn$_{3}$, as shown in Fig.~\ref{fig:USn3amptau}, yielding $\Delta \approx (90 \pm 20)$~K. The good fits show that the slow relaxation component in both UAl$_{3}$ and USn$_{3}$ can be described by assuming EHPs relaxing across the hybridization gap near the Fermi surface. More interestingly, the extracted hybridization gap in UAl$_{3}$ is larger than in USn$_{3}$, in qualitative agreement with the band structure results. This comparison of hybridization gap is also consistent with that of spin fluctuation energy scale $T_{sf}$ discussed above. Our results show that the ultrafast pump-probe technique is sensitive to the hybridization of \textit{f}-electron orbitals with the conduction electron orbitals.

\section{UG\lowercase{a}$_{3}$}
We now turn to the relaxation dynamics of UGa$_{3}$. UGa$_{3}$ is not a spin fluctuation system --- it is a SDW system with N\'{e}el temperature $T_{N}$=65~K. It is a moderate heavy fermion with Sommerfeld coefficient 52~mJ/K$^{2}$.mol,\cite{Biasini2005} and is reported to follow a modified Curie-Weiss law behavior \cite{Zhou1987} with $T^{\ast}$=2080~K which is indicative of strong hybridization in this compound. The 5\textit{f} electrons in UGa$_{3}$ can be considered itinerant because of the large hybridization of 5\textit{f} orbitals with conduction electron orbitals. The photoinduced change in reflectivity, as shown in Fig.~\ref{fig:UGa3}(a), can be fitted with a single exponential decay $\Delta R/R (t)= A(T) \exp (-t/\tau)$. The extracted relaxation amplitude $A(T)$ and time $\tau (T)$ are shown in Fig.~\ref{fig:UGa3}(b) and Fig.~\ref{fig:UGa3}(c), respectively. Upon entering the SDW phase, $A(T)$ increases with decreasing temperature. However, instead of monotonically increasing as in UAl$_{3}$ and USn$_{3}$, $A(T)$ now attains a maximum at $\sim$40~K and starts decreasing with decreasing temperature (see Fig.~\ref{fig:UGa3}(b)). Concurrently, $\tau$ exhibits a quasi-divergence at $T_{N}$, consistent with that observed in itinerant antiferromagnets UNiGa$_{5}$ and UPtGa$_{5}$, where the opening up of the SDW gap causes a bottleneck in quasiparticle relaxation.~\cite{Chia2006,Chia2011} In contrast to UNiGa$_{5}$ and UPtGa$_{5}$, however, where $\tau$ increases with decreasing temperature at low temperatures, $\tau$ in UGa$_{3}$ shows a (1) shoulder (or change in curvature) at 40~K, and (2) decrease with decreasing temperature. An anomaly at a spin-reorientation temperature $T_{sr}$=40~K has been reported in other measurements of UGa$_{3}$, whether in the presence of a magnetic field (nuclear magnetic resonance, neutron scattering, magnetic susceptibility),~\cite{Kaczorowski1998,Dervenagas1999,Kambe2002a,Kambe2003,Kambe2005a} or in the absence of a magnetic field (thermal conductivity and neutron scattering).~\cite{Kaczorowski1998,Dervenagas1999} This anomaly has been associated with a reorientation of the ordered magnetic moments, which induces strong modifications of the uranium 5\textit{f} orbitals.\cite{Kambe2003} The fact that the transition is observed in the absence of magnetic field is an indication that the bump we see at 40~K in our pump probe measurement is not an artifact, but corresponds to the moment reorientation as has been reported in other measurements mentioned above.

We use the model proposed by Kabanov \textit{et al.}~\cite{Kabanov99} to analyze the temperature dependence of $A$. The temperature-dependence of the relaxation amplitude in the SDW state for an isotropic temperature-\textit{dependent} gap $\Delta_{SDW}(T)$ is given by (writing $\Delta_{SDW}(T)$ as $\Delta(T)$)
\begin{equation}
A(T)=\frac{\epsilon_{I}/(\Delta(T)+k_{B}T/2)}{1+\zeta
\sqrt{\frac{2k_{B}T}{\pi\Delta(T)}}\exp[-\Delta(T)/k_{B}T]}, \label{eqn:amplitudea}
\end{equation} where $\epsilon_{I}$ the pump laser intensity per unit cell, $\zeta$ is a constant, and $\Delta(T)$ obeys a weak-coupling BCS temperature dependence. The above expression for $A(T)$ describes a reduction in the photoexcited quasiparticle density with increase in temperature, due to the decrease in gap energy and corresponding enhanced phonon emission during the initial relaxation. A good fit between the experimental $A(T)$ and Eq.~(\ref{eqn:amplitudea}) can only be made from $T_{N}$ down to $\sim$40~K, where $T_{N}$=55~K is a fitting parameter. In the SDW state ($T < T_{N}$), the temperature-dependence of $\tau$ cab be obtained from Eq.~(\ref{eqn:RTtau}), but can be written in the alternative form (writing $\Delta_{SDW}(T)$ as $\Delta(T)$) \cite{Nair2010}
\begin{multline}
\tau^{-1}(T)= \Gamma \{{\delta A(T)+ \eta \sqrt{\Delta(T)T} \exp[-\Delta(T)/T]}\}\\
 \times \left[\Delta(T) + \alpha T \Delta(T)^{4} \right].
\label{eqn:RTtauAlt}
\end{multline} The fit of $\tau (T)$ to Eq.~(\ref{eqn:RTtauAlt}) is shown in Fig.~\ref{fig:UGa3}(c). Once again, a good fit is obtained only from $T_{N}$ to $\sim$30~K, close to $T_{sr}$. Below $T_{sr}$, the fit deviates from the experimental data, consistent with the existence of another transition at $T_{sr}$.

\section{Conclusion}
We have performed time-resolved photoinduced change in reflectivity measurements on three isostructural uranium compounds --- UAl$_{3}$, UGa$_{3}$ and USn$_{3}$. The values of $T_{sf}$, a measure of the degree of hybridization, in UAl$_{3}$ and USn$_{3}$, are consistent with data from other measurements.  Our fit of the slow component to the Rothwarf-Taylor model shows that the slow component can be described by assuming electron-hole pairs relaxing across the hybridization gap. We have thus shown the pump probe technique to be sensitive to c-\textit{f} hybridization. Our data on UGa$_{3}$ is consistent with the formation of a SDW gap at $T_{N}$=60~K, and a reorientation of magnetic moments at $T_{sr}$=40~K.

\section{Acknowledgements}
E.E.M.C. acknowledges support from G. T. Seaborg Postdoctoral Fellowship, the Singapore Ministry of Education Academic Research Fund Tier 2 (ARC23/08), as well as the National Research Foundation Competitive Research Programme (NRF-CRP4-2008-04). Work at Los Alamos was supported by the U.S. DOE at LANL under Contract No. DE-AC52-06NA25396, the U.S. DOE Office of Basic Energy Sciences, and the LDRD Program at LANL. The electronic structure calculations were performed on a computer cluster at the Center for Integrated Nanotechnologies, a U.S. DOE Office of Basic Energy Sciences user facility.

\bibliography{Ultrafast,UX3}

\providecommand{\noopsort}[1]{}\providecommand{\singleletter}[1]{#1}%
\begin{thebibliography}{41}
\expandafter\ifx\csname natexlab\endcsname\relax\def\natexlab#1{#1}\fi
\expandafter\ifx\csname bibnamefont\endcsname\relax
  \def\bibnamefont#1{#1}\fi
\expandafter\ifx\csname bibfnamefont\endcsname\relax
  \def\bibfnamefont#1{#1}\fi
\expandafter\ifx\csname citenamefont\endcsname\relax
  \def\citenamefont#1{#1}\fi
\expandafter\ifx\csname url\endcsname\relax
  \def\url#1{\texttt{#1}}\fi
\expandafter\ifx\csname urlprefix\endcsname\relax\def\urlprefix{URL }\fi
\providecommand{\bibinfo}[2]{#2}
\providecommand{\eprint}[2][]{\url{#2}}

\bibitem[{\citenamefont{Fournier and Tro\'{c}}(1985)}]{Fournier1985}
\bibinfo{author}{\bibfnamefont{J.~M.} \bibnamefont{Fournier}} \bibnamefont{and}
  \bibinfo{author}{\bibfnamefont{R.}~\bibnamefont{Tro\'{c}}},
  \emph{\bibinfo{title}{{Handbook on the Physics and Chemistry of the
  Actinides}}}, vol.~\bibinfo{volume}{2} (\bibinfo{publisher}{North-Holland,
  Amsterdam}, \bibinfo{year}{1985}).

\bibitem[{\citenamefont{Hill}(1970)}]{Hill1970}
\bibinfo{author}{\bibfnamefont{H.~H.} \bibnamefont{Hill}}, in
  \emph{\bibinfo{booktitle}{Plutonium and other Actinides}}, edited by
  \bibinfo{editor}{\bibfnamefont{W.~N.} \bibnamefont{Miner}}
  (\bibinfo{publisher}{Nuclear Materials Sciences, AIME, New York},
  \bibinfo{year}{1970}), vol.~\bibinfo{volume}{17}, pp. \bibinfo{pages}{2--17}.

\bibitem[{\citenamefont{Koelling et~al.}(1985)\citenamefont{Koelling, Dunlap,
  and Crabtree}}]{Koelling1985}
\bibinfo{author}{\bibfnamefont{D.~D.} \bibnamefont{Koelling}},
  \bibinfo{author}{\bibfnamefont{B.~D.} \bibnamefont{Dunlap}},
  \bibnamefont{and} \bibinfo{author}{\bibfnamefont{G.~W.}
  \bibnamefont{Crabtree}}, \bibinfo{journal}{Phys. Rev. B}
  \textbf{\bibinfo{volume}{31}}, \bibinfo{pages}{4966} (\bibinfo{year}{1985}).

\bibitem[{\citenamefont{Ott et~al.}(1985)\citenamefont{Ott, Hulliger, Rudigier,
  and Fisk}}]{Ott1985}
\bibinfo{author}{\bibfnamefont{H.~R.} \bibnamefont{Ott}},
  \bibinfo{author}{\bibfnamefont{F.}~\bibnamefont{Hulliger}},
  \bibinfo{author}{\bibfnamefont{H.}~\bibnamefont{Rudigier}}, \bibnamefont{and}
  \bibinfo{author}{\bibfnamefont{Z.}~\bibnamefont{Fisk}},
  \bibinfo{journal}{Phys. Rev. B} \textbf{\bibinfo{volume}{31}},
  \bibinfo{pages}{1329} (\bibinfo{year}{1985}).

\bibitem[{\citenamefont{Arko et~al.}(1972)\citenamefont{Arko, Brodsky, and
  Nellis}}]{Arko1972}
\bibinfo{author}{\bibfnamefont{A.~J.} \bibnamefont{Arko}},
  \bibinfo{author}{\bibfnamefont{M.~B.} \bibnamefont{Brodsky}},
  \bibnamefont{and} \bibinfo{author}{\bibfnamefont{W.~J.}
  \bibnamefont{Nellis}}, \bibinfo{journal}{Phys. Rev. B}
  \textbf{\bibinfo{volume}{5}}, \bibinfo{pages}{4564} (\bibinfo{year}{1972}).

\bibitem[{\citenamefont{Jullien and Coqblin}(1976)}]{Jullien1976}
\bibinfo{author}{\bibfnamefont{R.}~\bibnamefont{Jullien}} \bibnamefont{and}
  \bibinfo{author}{\bibfnamefont{B.}~\bibnamefont{Coqblin}},
  \bibinfo{journal}{J. Low Temp. Phys.} \textbf{\bibinfo{volume}{22}},
  \bibinfo{pages}{437} (\bibinfo{year}{1976}).

\bibitem[{\citenamefont{Buschow and van Daal}(1972)}]{Buschow1972}
\bibinfo{author}{\bibfnamefont{K.~H.~J.} \bibnamefont{Buschow}}
  \bibnamefont{and} \bibinfo{author}{\bibfnamefont{H.~J.} \bibnamefont{van
  Daal}}, \bibinfo{journal}{AIP Conference Proceedings}
  \textbf{\bibinfo{volume}{5}}, \bibinfo{pages}{1464} (\bibinfo{year}{1972}).

\bibitem[{\citenamefont{Jullien et~al.}(1973)\citenamefont{Jullien,
  B{\'e}al-Monod, and Coqblin}}]{Jullien1973}
\bibinfo{author}{\bibfnamefont{R.}~\bibnamefont{Jullien}},
  \bibinfo{author}{\bibfnamefont{M.~T.} \bibnamefont{B{\'e}al-Monod}},
  \bibnamefont{and} \bibinfo{author}{\bibfnamefont{B.}~\bibnamefont{Coqblin}},
  \bibinfo{journal}{Phys. Rev. Lett.} \textbf{\bibinfo{volume}{30}},
  \bibinfo{pages}{1057} (\bibinfo{year}{1973}).

\bibitem[{\citenamefont{Jullien et~al.}(1974)\citenamefont{Jullien,
  B{\'e}al-Monod, and Coqblin}}]{Jullien1974a}
\bibinfo{author}{\bibfnamefont{R.}~\bibnamefont{Jullien}},
  \bibinfo{author}{\bibfnamefont{M.~T.} \bibnamefont{B{\'e}al-Monod}},
  \bibnamefont{and} \bibinfo{author}{\bibfnamefont{B.}~\bibnamefont{Coqblin}},
  \bibinfo{journal}{Phys. Rev. B} \textbf{\bibinfo{volume}{9}},
  \bibinfo{pages}{1441} (\bibinfo{year}{1974}).

\bibitem[{\citenamefont{Lin et~al.}(1985)\citenamefont{Lin, Zhou, Crow, and
  Guertin}}]{Lin1985}
\bibinfo{author}{\bibfnamefont{C.~L.} \bibnamefont{Lin}},
  \bibinfo{author}{\bibfnamefont{L.~W.} \bibnamefont{Zhou}},
  \bibinfo{author}{\bibfnamefont{J.~E.} \bibnamefont{Crow}}, \bibnamefont{and}
  \bibinfo{author}{\bibfnamefont{R.~P.} \bibnamefont{Guertin}},
  \bibinfo{journal}{J. Appl. Phys.} \textbf{\bibinfo{volume}{57}},
  \bibinfo{pages}{3146} (\bibinfo{year}{1985}).

\bibitem[{\citenamefont{Yuen et~al.}(1990)\citenamefont{Yuen, Gao, Perez, and
  Crow}}]{Yuen1990}
\bibinfo{author}{\bibfnamefont{T.}~\bibnamefont{Yuen}},
  \bibinfo{author}{\bibfnamefont{Y.}~\bibnamefont{Gao}},
  \bibinfo{author}{\bibfnamefont{I.}~\bibnamefont{Perez}}, \bibnamefont{and}
  \bibinfo{author}{\bibfnamefont{J.~E.} \bibnamefont{Crow}},
  \bibinfo{journal}{J. Appl. Phys.} \textbf{\bibinfo{volume}{67}},
  \bibinfo{pages}{4827} (\bibinfo{year}{1990}).

\bibitem[{\citenamefont{Chia et~al.}(2007)\citenamefont{Chia, Zhu, Talbayev,
  Averitt, Taylor, Oh, Jo, and Lee}}]{Chia2007}
\bibinfo{author}{\bibfnamefont{E.~E.~M.} \bibnamefont{Chia}},
  \bibinfo{author}{\bibfnamefont{J.-X.} \bibnamefont{Zhu}},
  \bibinfo{author}{\bibfnamefont{D.}~\bibnamefont{Talbayev}},
  \bibinfo{author}{\bibfnamefont{R.~D.} \bibnamefont{Averitt}},
  \bibinfo{author}{\bibfnamefont{A.~J.} \bibnamefont{Taylor}},
  \bibinfo{author}{\bibfnamefont{K.~H.} \bibnamefont{Oh}},
  \bibinfo{author}{\bibfnamefont{I.~S.} \bibnamefont{Jo}}, \bibnamefont{and}
  \bibinfo{author}{\bibfnamefont{S.~I.} \bibnamefont{Lee}},
  \bibinfo{journal}{Phys. Rev. Lett.} \textbf{\bibinfo{volume}{99}},
  \bibinfo{pages}{147008} (\bibinfo{year}{2007}).

\bibitem[{\citenamefont{Nair et~al.}(2010)\citenamefont{Nair, Zou, Chia, Zhu,
  Panagopoulos, Ishida, and Uchida}}]{Nair2010}
\bibinfo{author}{\bibfnamefont{S.~K.} \bibnamefont{Nair}},
  \bibinfo{author}{\bibfnamefont{X.~Q.} \bibnamefont{Zou}},
  \bibinfo{author}{\bibfnamefont{E.~E.~M.} \bibnamefont{Chia}},
  \bibinfo{author}{\bibfnamefont{J.-X.} \bibnamefont{Zhu}},
  \bibinfo{author}{\bibfnamefont{C.}~\bibnamefont{Panagopoulos}},
  \bibinfo{author}{\bibfnamefont{S.}~\bibnamefont{Ishida}}, \bibnamefont{and}
  \bibinfo{author}{\bibfnamefont{S.}~\bibnamefont{Uchida}},
  \bibinfo{journal}{Phys. Rev. B} \textbf{\bibinfo{volume}{82}},
  \bibinfo{pages}{212503} (\bibinfo{year}{2010}).

\bibitem[{\citenamefont{Chia et~al.}(2006)\citenamefont{Chia, Zhu, Lee, Hur,
  Moreno, Bauer, Durakiewicz, Averitt, Sarrao, and Taylor}}]{Chia2006}
\bibinfo{author}{\bibfnamefont{E.~E.~M.} \bibnamefont{Chia}},
  \bibinfo{author}{\bibfnamefont{J.-X.} \bibnamefont{Zhu}},
  \bibinfo{author}{\bibfnamefont{H.~J.} \bibnamefont{Lee}},
  \bibinfo{author}{\bibfnamefont{N.}~\bibnamefont{Hur}},
  \bibinfo{author}{\bibfnamefont{N.~O.} \bibnamefont{Moreno}},
  \bibinfo{author}{\bibfnamefont{E.~D.} \bibnamefont{Bauer}},
  \bibinfo{author}{\bibfnamefont{T.}~\bibnamefont{Durakiewicz}},
  \bibinfo{author}{\bibfnamefont{R.~D.} \bibnamefont{Averitt}},
  \bibinfo{author}{\bibfnamefont{J.~L.} \bibnamefont{Sarrao}},
  \bibnamefont{and} \bibinfo{author}{\bibfnamefont{A.~J.}
  \bibnamefont{Taylor}}, \bibinfo{journal}{Phys. Rev. B}
  \textbf{\bibinfo{volume}{74}}, \bibinfo{pages}{140409}
  (\bibinfo{year}{2006}).

\bibitem[{\citenamefont{Chia et~al.}(2011)\citenamefont{Chia, Zhu, Talbayev,
  Lee, N.Hur, Moreno, Averitt, Sarrao, and Taylor}}]{Chia2011}
\bibinfo{author}{\bibfnamefont{E.~E.~M.} \bibnamefont{Chia}},
  \bibinfo{author}{\bibfnamefont{J.-X.} \bibnamefont{Zhu}},
  \bibinfo{author}{\bibfnamefont{D.}~\bibnamefont{Talbayev}},
  \bibinfo{author}{\bibfnamefont{H.~J.} \bibnamefont{Lee}},
  \bibinfo{author}{\bibnamefont{N.Hur}}, \bibinfo{author}{\bibfnamefont{N.~O.}
  \bibnamefont{Moreno}}, \bibinfo{author}{\bibfnamefont{R.~D.}
  \bibnamefont{Averitt}}, \bibinfo{author}{\bibfnamefont{J.~L.}
  \bibnamefont{Sarrao}}, \bibnamefont{and}
  \bibinfo{author}{\bibfnamefont{A.~J.} \bibnamefont{Taylor}},
  \bibinfo{journal}{Phys. Rev. B} \textbf{\bibinfo{volume}{84}},
  \bibinfo{pages}{174412} (\bibinfo{year}{2011}).

\bibitem[{\citenamefont{Talbayev et~al.}(2010)\citenamefont{Talbayev, Burch,
  Chia, Trugman, Zhu, Bauer, Kennison, Mitchell, Thompson, Sarrao
  et~al.}}]{Talbayev10}
\bibinfo{author}{\bibfnamefont{D.}~\bibnamefont{Talbayev}},
  \bibinfo{author}{\bibfnamefont{K.~S.} \bibnamefont{Burch}},
  \bibinfo{author}{\bibfnamefont{E.~E.~M.} \bibnamefont{Chia}},
  \bibinfo{author}{\bibfnamefont{S.~A.} \bibnamefont{Trugman}},
  \bibinfo{author}{\bibfnamefont{J.-X.} \bibnamefont{Zhu}},
  \bibinfo{author}{\bibfnamefont{E.~D.} \bibnamefont{Bauer}},
  \bibinfo{author}{\bibfnamefont{J.~A.} \bibnamefont{Kennison}},
  \bibinfo{author}{\bibfnamefont{J.~N.} \bibnamefont{Mitchell}},
  \bibinfo{author}{\bibfnamefont{J.}~\bibnamefont{Thompson}},
  \bibinfo{author}{\bibfnamefont{J.~L.} \bibnamefont{Sarrao}},
  \bibnamefont{et~al.}, \bibinfo{journal}{Phys. Rev. Lett.}
  \textbf{\bibinfo{volume}{104}}, \bibinfo{pages}{227002}
  (\bibinfo{year}{2010}).

\bibitem[{\citenamefont{Demsar et~al.}(2006)\citenamefont{Demsar, Sarrao, and
  Taylor}}]{Demsar06b}
\bibinfo{author}{\bibfnamefont{J.}~\bibnamefont{Demsar}},
  \bibinfo{author}{\bibfnamefont{J.~L.} \bibnamefont{Sarrao}},
  \bibnamefont{and} \bibinfo{author}{\bibfnamefont{A.~J.}
  \bibnamefont{Taylor}}, \bibinfo{journal}{J. Phys.: Condens. Matter}
  \textbf{\bibinfo{volume}{18}}, \bibinfo{pages}{R281} (\bibinfo{year}{2006}).

\bibitem[{\citenamefont{Van~Maaren et~al.}(1974)\citenamefont{Van~Maaren,
  Van~Daal, Buschow, and Schinkel}}]{VanMaaren1974}
\bibinfo{author}{\bibfnamefont{M.~H.} \bibnamefont{Van~Maaren}},
  \bibinfo{author}{\bibfnamefont{H.~J.} \bibnamefont{Van~Daal}},
  \bibinfo{author}{\bibfnamefont{K.~H.~J.} \bibnamefont{Buschow}},
  \bibnamefont{and} \bibinfo{author}{\bibfnamefont{C.~J.}
  \bibnamefont{Schinkel}}, \bibinfo{journal}{Solid State Commun.}
  \textbf{\bibinfo{volume}{14}}, \bibinfo{pages}{145} (\bibinfo{year}{1974}).

\bibitem[{\citenamefont{Lupsa et~al.}(1994)\citenamefont{Lupsa, Lucaci, and
  Burzo}}]{Lupsa1994}
\bibinfo{author}{\bibfnamefont{I.}~\bibnamefont{Lupsa}},
  \bibinfo{author}{\bibfnamefont{P.}~\bibnamefont{Lucaci}}, \bibnamefont{and}
  \bibinfo{author}{\bibfnamefont{E.}~\bibnamefont{Burzo}}, \bibinfo{journal}{J.
  Alloys Compd.} \textbf{\bibinfo{volume}{204}}, \bibinfo{pages}{247 }
  (\bibinfo{year}{1994}).

\bibitem[{\citenamefont{Lupsa et~al.}(1996)\citenamefont{Lupsa, Burzo, and
  Lucaci}}]{Lupsa1996}
\bibinfo{author}{\bibfnamefont{I.}~\bibnamefont{Lupsa}},
  \bibinfo{author}{\bibfnamefont{E.}~\bibnamefont{Burzo}}, \bibnamefont{and}
  \bibinfo{author}{\bibfnamefont{P.}~\bibnamefont{Lucaci}},
  \bibinfo{journal}{J. Magn. Magn. Mater.} \textbf{\bibinfo{volume}{157}},
  \bibinfo{pages}{696} (\bibinfo{year}{1996}).

\bibitem[{\citenamefont{Norman et~al.}(1986)\citenamefont{Norman, Bader, and
  Kierstead}}]{Norman1986}
\bibinfo{author}{\bibfnamefont{M.~R.} \bibnamefont{Norman}},
  \bibinfo{author}{\bibfnamefont{S.~D.} \bibnamefont{Bader}}, \bibnamefont{and}
  \bibinfo{author}{\bibfnamefont{H.~A.} \bibnamefont{Kierstead}},
  \bibinfo{journal}{Phys. Rev. B} \textbf{\bibinfo{volume}{33}},
  \bibinfo{pages}{8035} (\bibinfo{year}{1986}).

\bibitem[{\citenamefont{Cornelius et~al.}(1999)\citenamefont{Cornelius, Arko,
  Sarrao, Thompson, Hundley, Booth, Harrison, and Oppeneer}}]{Cornelius1999}
\bibinfo{author}{\bibfnamefont{A.~L.} \bibnamefont{Cornelius}},
  \bibinfo{author}{\bibfnamefont{A.~J.} \bibnamefont{Arko}},
  \bibinfo{author}{\bibfnamefont{J.~L.} \bibnamefont{Sarrao}},
  \bibinfo{author}{\bibfnamefont{J.~D.} \bibnamefont{Thompson}},
  \bibinfo{author}{\bibfnamefont{M.~F.} \bibnamefont{Hundley}},
  \bibinfo{author}{\bibfnamefont{C.~H.} \bibnamefont{Booth}},
  \bibinfo{author}{\bibfnamefont{N.}~\bibnamefont{Harrison}}, \bibnamefont{and}
  \bibinfo{author}{\bibfnamefont{P.~M.} \bibnamefont{Oppeneer}},
  \bibinfo{journal}{Phys. Rev. B} \textbf{\bibinfo{volume}{59}},
  \bibinfo{pages}{14473} (\bibinfo{year}{1999}).

\bibitem[{\citenamefont{Aoki et~al.}(2000)\citenamefont{Aoki, Watanabe, Inada,
  Settai, Sugiyama, Harima, Inoue, Kindo, Yamamoto, Haga et~al.}}]{Aoki2000}
\bibinfo{author}{\bibfnamefont{D.}~\bibnamefont{Aoki}},
  \bibinfo{author}{\bibfnamefont{N.}~\bibnamefont{Watanabe}},
  \bibinfo{author}{\bibfnamefont{Y.}~\bibnamefont{Inada}},
  \bibinfo{author}{\bibfnamefont{R.}~\bibnamefont{Settai}},
  \bibinfo{author}{\bibfnamefont{K.}~\bibnamefont{Sugiyama}},
  \bibinfo{author}{\bibfnamefont{H.}~\bibnamefont{Harima}},
  \bibinfo{author}{\bibfnamefont{T.}~\bibnamefont{Inoue}},
  \bibinfo{author}{\bibfnamefont{K.}~\bibnamefont{Kindo}},
  \bibinfo{author}{\bibfnamefont{E.}~\bibnamefont{Yamamoto}},
  \bibinfo{author}{\bibfnamefont{Y.}~\bibnamefont{Haga}}, \bibnamefont{et~al.},
  \bibinfo{journal}{J. Phys. Soc. Jpn.} \textbf{\bibinfo{volume}{69}},
  \bibinfo{pages}{2609} (\bibinfo{year}{2000}).

\bibitem[{\citenamefont{Loewenhaupt and Loong}(1990)}]{Loewenhaupt1990}
\bibinfo{author}{\bibfnamefont{M.}~\bibnamefont{Loewenhaupt}} \bibnamefont{and}
  \bibinfo{author}{\bibfnamefont{C.~K.} \bibnamefont{Loong}},
  \bibinfo{journal}{Phys. Rev. B} \textbf{\bibinfo{volume}{41}},
  \bibinfo{pages}{9294} (\bibinfo{year}{1990}).

\bibitem[{\citenamefont{Sugiyama et~al.}(2002)\citenamefont{Sugiyama, Iizuka
  et~al.}}]{Sugiyama2002}
\bibinfo{author}{\bibfnamefont{K.}~\bibnamefont{Sugiyama}},
  \bibinfo{author}{\bibfnamefont{T.}~\bibnamefont{Iizuka}},
  \bibnamefont{et~al.}, \bibinfo{journal}{J. Phys. Soc. Jpn.}
  \textbf{\bibinfo{volume}{71}}, \bibinfo{pages}{326} (\bibinfo{year}{2002}),
  ISSN \bibinfo{issn}{0031-9015}.

\bibitem[{\citenamefont{Blaha et~al.}(2001)\citenamefont{Blaha, Schwarz,
  Madsen, Kvasnicka, and Luitz}}]{Blaha01}
\bibinfo{author}{\bibfnamefont{P.}~\bibnamefont{Blaha}},
  \bibinfo{author}{\bibfnamefont{K.}~\bibnamefont{Schwarz}},
  \bibinfo{author}{\bibfnamefont{G.~K.~H.} \bibnamefont{Madsen}},
  \bibinfo{author}{\bibfnamefont{D.}~\bibnamefont{Kvasnicka}},
  \bibnamefont{and} \bibinfo{author}{\bibfnamefont{J.}~\bibnamefont{Luitz}},
  \emph{\bibinfo{title}{WIEN2k, An Augmented Plane Wave Plus Local Orbitals
  Program for CalculatingCrystal Properties}} (\bibinfo{publisher}{Vienna
  University of Technology, Austria}, \bibinfo{year}{2001}).

\bibitem[{\citenamefont{Perdew et~al.}(1996)\citenamefont{Perdew, Burke, and
  Ernzerhof}}]{Perdew1996}
\bibinfo{author}{\bibfnamefont{J.~P.} \bibnamefont{Perdew}},
  \bibinfo{author}{\bibfnamefont{K.}~\bibnamefont{Burke}}, \bibnamefont{and}
  \bibinfo{author}{\bibfnamefont{M.}~\bibnamefont{Ernzerhof}},
  \bibinfo{journal}{Phys. Rev. Lett.} \textbf{\bibinfo{volume}{77}},
  \bibinfo{pages}{3865} (\bibinfo{year}{1996}).

\bibitem[{\citenamefont{Zhu et~al.}(2012)\citenamefont{Zhu, Tobash, Bauer,
  Ronning, Scott, Haule, Kotliar, Albers, and Wills}}]{Zhu2012}
\bibinfo{author}{\bibfnamefont{J.-X.} \bibnamefont{Zhu}},
  \bibinfo{author}{\bibfnamefont{P.~H.} \bibnamefont{Tobash}},
  \bibinfo{author}{\bibfnamefont{E.~D.} \bibnamefont{Bauer}},
  \bibinfo{author}{\bibfnamefont{F.}~\bibnamefont{Ronning}},
  \bibinfo{author}{\bibfnamefont{B.~L.} \bibnamefont{Scott}},
  \bibinfo{author}{\bibfnamefont{K.}~\bibnamefont{Haule}},
  \bibinfo{author}{\bibfnamefont{G.}~\bibnamefont{Kotliar}},
  \bibinfo{author}{\bibfnamefont{R.~C.} \bibnamefont{Albers}},
  \bibnamefont{and} \bibinfo{author}{\bibfnamefont{J.~M.} \bibnamefont{Wills}},
  \bibinfo{journal}{Europhys. Lett.} \textbf{\bibinfo{volume}{97}},
  \bibinfo{pages}{57001} (\bibinfo{year}{2012}).

\bibitem[{\citenamefont{Lobad et~al.}(2000)\citenamefont{Lobad, Taylor, Sarrao,
  and Trugman}}]{Lobad2000}
\bibinfo{author}{\bibfnamefont{A.~I.} \bibnamefont{Lobad}},
  \bibinfo{author}{\bibfnamefont{A.~J.} \bibnamefont{Taylor}},
  \bibinfo{author}{\bibfnamefont{J.~L.} \bibnamefont{Sarrao}},
  \bibnamefont{and} \bibinfo{author}{\bibfnamefont{S.~A.}
  \bibnamefont{Trugman}}, in \emph{\bibinfo{booktitle}{Quantum Electronics and
  Laser Science Conference, 2000. (QELS 2000). TechnicalDigest}}
  (\bibinfo{year}{2000}), pp. \bibinfo{pages}{159 -- 160}.

\bibitem[{\citenamefont{Arko and Koelling}(1978)}]{Arko1978}
\bibinfo{author}{\bibfnamefont{A.~J.} \bibnamefont{Arko}} \bibnamefont{and}
  \bibinfo{author}{\bibfnamefont{D.~D.} \bibnamefont{Koelling}},
  \bibinfo{journal}{Phys. Rev. B} \textbf{\bibinfo{volume}{17}},
  \bibinfo{pages}{3104} (\bibinfo{year}{1978}).

\bibitem[{\citenamefont{Rothwarf and Taylor}(1967)}]{Rothwarf67}
\bibinfo{author}{\bibfnamefont{A.}~\bibnamefont{Rothwarf}} \bibnamefont{and}
  \bibinfo{author}{\bibfnamefont{B.~N.} \bibnamefont{Taylor}},
  \bibinfo{journal}{Phys. Rev. Lett.} \textbf{\bibinfo{volume}{19}},
  \bibinfo{pages}{27} (\bibinfo{year}{1967}).

\bibitem[{\citenamefont{Demsar et~al.}(2003)\citenamefont{Demsar, Averitt,
  Taylor, Kabanov, N.Kang, Kim, Choi, and Lee}}]{Demsar03c}
\bibinfo{author}{\bibfnamefont{J.}~\bibnamefont{Demsar}},
  \bibinfo{author}{\bibfnamefont{R.~D.} \bibnamefont{Averitt}},
  \bibinfo{author}{\bibfnamefont{A.~J.} \bibnamefont{Taylor}},
  \bibinfo{author}{\bibfnamefont{V.~V.} \bibnamefont{Kabanov}},
  \bibinfo{author}{\bibfnamefont{W.}~\bibnamefont{N.Kang}},
  \bibinfo{author}{\bibfnamefont{H.~J.} \bibnamefont{Kim}},
  \bibinfo{author}{\bibfnamefont{E.~M.} \bibnamefont{Choi}}, \bibnamefont{and}
  \bibinfo{author}{\bibfnamefont{S.~I.} \bibnamefont{Lee}},
  \bibinfo{journal}{Phys. Rev. Lett.} \textbf{\bibinfo{volume}{91}},
  \bibinfo{pages}{267002} (\bibinfo{year}{2003}).

\bibitem[{\citenamefont{Kabanov et~al.}(2005)\citenamefont{Kabanov, Demsar, and
  Mihailovic}}]{Kabanov05}
\bibinfo{author}{\bibfnamefont{V.~V.} \bibnamefont{Kabanov}},
  \bibinfo{author}{\bibfnamefont{J.}~\bibnamefont{Demsar}}, \bibnamefont{and}
  \bibinfo{author}{\bibfnamefont{D.}~\bibnamefont{Mihailovic}},
  \bibinfo{journal}{Phys. Rev. Lett.} \textbf{\bibinfo{volume}{95}},
  \bibinfo{pages}{147002} (\bibinfo{year}{2005}).

\bibitem[{\citenamefont{Biasini et~al.}(2005)\citenamefont{Biasini, Rusz,
  Ferro, and Czopnik}}]{Biasini2005}
\bibinfo{author}{\bibfnamefont{M.}~\bibnamefont{Biasini}},
  \bibinfo{author}{\bibfnamefont{J.}~\bibnamefont{Rusz}},
  \bibinfo{author}{\bibfnamefont{G.}~\bibnamefont{Ferro}}, \bibnamefont{and}
  \bibinfo{author}{\bibfnamefont{A.}~\bibnamefont{Czopnik}},
  \bibinfo{journal}{Acta Phys. Pol. A} \textbf{\bibinfo{volume}{107}},
  \bibinfo{pages}{554} (\bibinfo{year}{2005}).

\bibitem[{\citenamefont{Zhou et~al.}(1987)\citenamefont{Zhou, Jee, Lin, Crow,
  Bloom, and Guertin}}]{Zhou1987}
\bibinfo{author}{\bibfnamefont{L.~W.} \bibnamefont{Zhou}},
  \bibinfo{author}{\bibfnamefont{C.~S.} \bibnamefont{Jee}},
  \bibinfo{author}{\bibfnamefont{C.~L.} \bibnamefont{Lin}},
  \bibinfo{author}{\bibfnamefont{J.~E.} \bibnamefont{Crow}},
  \bibinfo{author}{\bibfnamefont{S.}~\bibnamefont{Bloom}}, \bibnamefont{and}
  \bibinfo{author}{\bibfnamefont{R.~P.} \bibnamefont{Guertin}},
  \bibinfo{journal}{J. Appl. Phys.} \textbf{\bibinfo{volume}{61}},
  \bibinfo{pages}{3377} (\bibinfo{year}{1987}).

\bibitem[{\citenamefont{Kaczorowski et~al.}(1998)\citenamefont{Kaczorowski,
  Klamut, Czopnik, and Je\.{z}owski}}]{Kaczorowski1998}
\bibinfo{author}{\bibfnamefont{D.}~\bibnamefont{Kaczorowski}},
  \bibinfo{author}{\bibfnamefont{P.~W.} \bibnamefont{Klamut}},
  \bibinfo{author}{\bibfnamefont{A.}~\bibnamefont{Czopnik}}, \bibnamefont{and}
  \bibinfo{author}{\bibfnamefont{A.}~\bibnamefont{Je\.{z}owski}},
  \bibinfo{journal}{J. Magn. Magn. Mater.} \textbf{\bibinfo{volume}{177}},
  \bibinfo{pages}{41} (\bibinfo{year}{1998}).

\bibitem[{\citenamefont{Dervenagas et~al.}(1999)\citenamefont{Dervenagas,
  Kaczorowski, Bourdarot, Burlet, Czopnik, and Lander}}]{Dervenagas1999}
\bibinfo{author}{\bibfnamefont{P.}~\bibnamefont{Dervenagas}},
  \bibinfo{author}{\bibfnamefont{D.}~\bibnamefont{Kaczorowski}},
  \bibinfo{author}{\bibfnamefont{F.}~\bibnamefont{Bourdarot}},
  \bibinfo{author}{\bibfnamefont{P.}~\bibnamefont{Burlet}},
  \bibinfo{author}{\bibfnamefont{A.}~\bibnamefont{Czopnik}}, \bibnamefont{and}
  \bibinfo{author}{\bibfnamefont{G.~H.} \bibnamefont{Lander}},
  \bibinfo{journal}{Physica B} \textbf{\bibinfo{volume}{269}},
  \bibinfo{pages}{368} (\bibinfo{year}{1999}).

\bibitem[{\citenamefont{Kambe et~al.}(2002)\citenamefont{Kambe, Kato, Sakai,
  Walstedt, Haga, Aoki, and {\^O}nuki}}]{Kambe2002a}
\bibinfo{author}{\bibfnamefont{S.}~\bibnamefont{Kambe}},
  \bibinfo{author}{\bibfnamefont{H.}~\bibnamefont{Kato}},
  \bibinfo{author}{\bibfnamefont{H.}~\bibnamefont{Sakai}},
  \bibinfo{author}{\bibfnamefont{R.~E.} \bibnamefont{Walstedt}},
  \bibinfo{author}{\bibfnamefont{Y.}~\bibnamefont{Haga}},
  \bibinfo{author}{\bibfnamefont{D.}~\bibnamefont{Aoki}}, \bibnamefont{and}
  \bibinfo{author}{\bibfnamefont{Y.}~\bibnamefont{{\^O}nuki}},
  \bibinfo{journal}{Physica B} \textbf{\bibinfo{volume}{312}},
  \bibinfo{pages}{902} (\bibinfo{year}{2002}).

\bibitem[{\citenamefont{Kambe et~al.}(2003)\citenamefont{Kambe, Kato, Sakai,
  Tokunaga, Walstedt, Haga, Yasuoka, and Aoki}}]{Kambe2003}
\bibinfo{author}{\bibfnamefont{S.}~\bibnamefont{Kambe}},
  \bibinfo{author}{\bibfnamefont{H.}~\bibnamefont{Kato}},
  \bibinfo{author}{\bibfnamefont{H.}~\bibnamefont{Sakai}},
  \bibinfo{author}{\bibfnamefont{Y.}~\bibnamefont{Tokunaga}},
  \bibinfo{author}{\bibfnamefont{R.~E.} \bibnamefont{Walstedt}},
  \bibinfo{author}{\bibfnamefont{Y.}~\bibnamefont{Haga}},
  \bibinfo{author}{\bibfnamefont{H.}~\bibnamefont{Yasuoka}}, \bibnamefont{and}
  \bibinfo{author}{\bibfnamefont{D.}~\bibnamefont{Aoki}},
  \bibinfo{journal}{Physica B} \textbf{\bibinfo{volume}{329}},
  \bibinfo{pages}{614} (\bibinfo{year}{2003}).

\bibitem[{\citenamefont{Kambe et~al.}(2005)\citenamefont{Kambe, Walstedt,
  Sakai, Tokunaga, Matsuda, Haga, and {\^O}nuki}}]{Kambe2005a}
\bibinfo{author}{\bibfnamefont{S.}~\bibnamefont{Kambe}},
  \bibinfo{author}{\bibfnamefont{R.~E.} \bibnamefont{Walstedt}},
  \bibinfo{author}{\bibfnamefont{H.}~\bibnamefont{Sakai}},
  \bibinfo{author}{\bibfnamefont{Y.}~\bibnamefont{Tokunaga}},
  \bibinfo{author}{\bibfnamefont{T.~D.} \bibnamefont{Matsuda}},
  \bibinfo{author}{\bibfnamefont{Y.}~\bibnamefont{Haga}}, \bibnamefont{and}
  \bibinfo{author}{\bibfnamefont{Y.}~\bibnamefont{{\^O}nuki}},
  \bibinfo{journal}{Phys. Rev. B} \textbf{\bibinfo{volume}{72}},
  \bibinfo{pages}{184437} (\bibinfo{year}{2005}).

\bibitem[{\citenamefont{Kabanov et~al.}(1999)\citenamefont{Kabanov, Demsar,
  Podobnik, and Mihailovic}}]{Kabanov99}
\bibinfo{author}{\bibfnamefont{V.~V.} \bibnamefont{Kabanov}},
  \bibinfo{author}{\bibfnamefont{J.}~\bibnamefont{Demsar}},
  \bibinfo{author}{\bibfnamefont{B.}~\bibnamefont{Podobnik}}, \bibnamefont{and}
  \bibinfo{author}{\bibfnamefont{D.}~\bibnamefont{Mihailovic}},
  \bibinfo{journal}{Phys. Rev. B} \textbf{\bibinfo{volume}{59}},
  \bibinfo{pages}{1497} (\bibinfo{year}{1999}).

\end{thebibliography}

\bigskip

\end{document}